\documentclass{aa}
\usepackage{psfig}
\usepackage{lscape}
\usepackage{amsmath}
\usepackage{amssymb}
\usepackage[figuresleft]{rotating}

\newcommand{\MC}{\multicolumn}
\newcommand{\kms}{km~s$^\mathrm{-1}$}
\newcommand{\sunn}{$_{\odot}$}
\newcounter{qub}
\begin{document}

\title{
Study of DDO~68:  nearest candidate for a young galaxy?
}

\author{Simon A. Pustilnik\inst{1}  \and
Alexei Y. Kniazev\inst{2,1} \and
Alexander G. Pramskij\inst{1}
}

\offprints{S. Pustilnik, \email{sap@sao.ru}}

\institute{
Special Astrophysical Observatory RAS, Nizhnij Arkhyz,
Karachai-Circassia,  369167 Russia
\and European Southern Observatory, Garching-bei-Munchen, Germany
}

 \date{Received March 21, 2005; accepted July 8, 2005}

\abstract{
We present the results of optical spectroscopy and imaging with the SAO 6\,m
telescope for the dwarf galaxy DDO~68 (UGC~5340=VV~542), falling into the
region of very low density of luminous (L $>$ L$_{*}$) galaxies
(Lynx-Cancer void).
Its deep images in $V,R$ bands and in the narrow H$\alpha$-filter show that
this galaxy has the very irregular morphology, with a long curved
tail on the South and a ring-like structure at the Northern edge. The latter
consists of 5 separate regions, in three of which we could measure O/H by the
classical T$_e$ method. Their weighted mean oxygen abundance corresponds to
12+$\log$(O/H)=7.21$\pm$0.03, coincident within uncertainties with those for
I~Zw~18.
The $(V-R)$ colour of DDO~68 is rather blue all over the galaxy,
indicating the youth of its stellar populations. Comparing the
$(V-R)^{0}$ colour of the underlying exponential disk of 0\fm12$\pm$0.04
with the PEGASE.2 models for the evolving stellar
clusters, we give the first estimate of the ages of the oldest stellar
population, which needs confirmation by the other colours and
the photometry of resolved stars.
These ages are in the range of 200--900 Myr for continuous star
formation law, and $\sim$100--115 Myr  for the instantaneous starburst.
We discuss the properties and the possible youth of this nearby object
($\sim$2.3 times closer than the famous young galaxy I~Zw~18) in the context
of its atypical environment.
   }

\authorrunning{S.A. Pustilnik et al.}

\titlerunning{Study DDO~68: nearest candidate for a young galaxy?}

   \maketitle

\section{Introduction}
\label{intro}

Some of dwarf gas-rich galaxies with very low metallicity were considered
as probable young galaxies since the seminal paper by Searle \& Sargent
(\cite{SS72}). The most metal-poor dwarf I~Zw~18 (with the value of
12+$\log$(O/H)=7.17--7.21, e.g., Alloin et al. \cite{Alloin},
Izotov et al. \cite{Izotov99}) has for 30 years been the Candidate No.1
of the local young galaxies. Recently, Izotov \& Thuan (\cite{IZw18CMD})
have demonstrated from the very deep Hubble Space Telescope (HST)
Colour-Magnitude Diagram (CMD) data that there are no stars in this object
older than 0.5 Gyr; that is, this nearby object (D=15 Mpc) is indeed a
genuine young local galaxy.

There are several other eXtremely Metal-Deficient objects (XMD, conditionally
with
the characteristic value Z$<$1/10~Z\sunn\footnote{The solar oxygen abundance
is accepted as 12+$\log$(O/H) = 8.66 according to Asplund et al.
(\cite{Solar04}).}, or 12+$\log$(O/H)$\leq$7.65; e.g., Kunth \& \"Ostlin
\cite{Kunth2000}), considered as good candidates to young local galaxies.
First of all, these are the components of the galaxy
pair SBS 0335--052 E and W, with 12+$\log$(O/H)=7.29 and 7.12, respectively
(e.g., Izotov et al. \cite{Izotov97}, Lipovetsky et al. \cite{Lipovetsky99},
Pustilnik et al. \cite{VLA,BTA}, Izotov \& Thuan \cite{IT05W}).
This unique pair at $D \sim$54 Mpc is, however,  3.6 times more distant
than I~Zw~18.

In our discussion, we distinguish truly young galaxies (similar to
I~Zw~18), in which their young stellar generation is made up of pregalactic
matter, from
the so-called tidal young dwarfs (e.g., Duc \& Mirabel \cite{Tidal}),
similar to Holmberg IX in the vicinity of M81.
Holmberg IX is devoid of old Red Giant Branch (RGB) stars (Makarova et al.
\cite{HoIX}), so its star formation began recently. But its gas metal content
is close to that of the massive parent galaxy ISM, from which it was recently
formed, so in this context such tidal dwarfs should be treated  not
as young but as rejuvenated.

It is worth noting that the very low ISM metallicity is considered only as an
indication of the possible unevolved status of a galaxy. In particular, from
the theoretical point of view, several evolutionary scenarios can lead to such
a result. Briefly, they include: a) the significant metal loss due to galactic
superwinds in the shallow gravitational wells of dwarf galaxies; b) the inflow
and the mixture of an unevolved intergalactic gas to an evolved galaxy; c)
the very slow astration and related ISM enrichment characteristic of the very
low surface brightness (LSB) galaxies, and d) the recent onset of the first
star formation episode. All but one correspond to an aged galaxy.
Moreover,
observationally among 16 known XMD galaxies at  distances less than
15 Mpc,  at least nine objects, either from the CMD analysis (e.g., Sextans A
and B, GR 8, Leo A) or from the red colours of the unresolved stellar
population in their outer regions (e.g., UGCA 20, UGC 2684, van Zee et al.
\cite{vZee_UA20,vZee_colors}), are recognized as  old systems.
Only one, I~Zw~18,  is now considered as a truly young object.

In searches for new XMD galaxies, we found that some of them populate the
`void' regions, where the distances from the void centers to luminous
galaxies (L $>$ L$_{*}$, L$_{*}$ correspond to M$_{\rm B}$=--19.6 for
H=72~\kms~Mpc$^{-1}$, assumed in the paper) exceed 4--5 Mpc for the smallest
voids and 8--10 Mpc for the more typical voids. In particular, the XMD blue
compact dwarf (BCD)  galaxy HS 0822+3542 with 12+$\log$(O/H)=7.38, appeared
near the centre of the nearby small Lynx-Cancer void (Kniazev et al.
\cite{Kniazev00}, Pustilnik et al. \cite{P_SAO0822}). Voids delineated by
luminous, massive galaxies are not
absolutely empty. Some numbers of the lower-mass galaxies fill in voids
(e.g., Salzer \cite{Salzer89}, Pustilnik et al. \cite{PULTG}, Lindner et al.
\cite{void_gal}, Popescu et al. \cite{Popescu97}, Grogin \& Geller
\cite{Grogin00} among others), probably forming the substructures of the
filament types (Lindner et al. \cite{void_gal}, Gottl\"ober et al.
\cite{Gott03}, Fairall et al. \cite{Fairall04}).
The very low-density environment
in voids and the scaled-down mass spectrum of the preformed DM halos,
coupled with the effect of reionization, are all expected to cause formation
of a somewhat different galaxy population from any in  more typical,
denser environments (e.g., Peebles \cite{Peebles01}, Gottl\"ober et al.
\cite{Gott03}).
A significantly reduced probability of galaxy interactions in voids, which is
suggested as an important factor of star formation history (e.g., Rojas
et al. \cite{SDSS_void}), can cause the differences in the rate of chemical
enrichment.

Some indication of the less evolved status of low-mass
galaxies in voids was presented, e.g., by Huchtmeier et al. (\cite{void_HI})
and Pustilnik et al. (\cite{HI_void}), based on \ion{H}{i} data. The higher
star formation (SF) activity of void galaxies in respect to `wall' galaxies,
detected by
Grogin \& Geller (\cite{Grogin00}) and Rojas et al. (\cite{SDSS_void}), also
indicates their larger available gas fuel reservoir. However, the metallicity
issue of the void galaxy population has not yet been properly addressed, mainly
due to observational difficulties.
Only the study of absorption lines  in Ly$\alpha$ clouds with  low column
densities in various environments led to the conclusion that such objects in
voids show a significantly lower content of heavy elements (Lu et al.
\cite{Lu98}).

Since the Lynx-Cancer void is one of the nearest (D$_{\rm center} \sim$11
Mpc), many of the underluminous
galaxies falling in this region are sufficiently bright,
so that in contrast to the situation in more distant voids, even inconspicuous
\ion{H}{ii} regions in more or less typical late-type dwarfs are
available for determining chemical abundance in `routine' programs.
Therefore, it was tempting to search in this void for some other XMD objects,
including likely young galaxy candidates.
To that end we performed the spectrophotometry of ten dwarfs in this region,
and the data will be presented separately elsewhere.
In this paper we report the discovery of an extremely low oxygen abundance
in dIm/BCD galaxy DDO~68 (UGC~5340=VV~542, J2000 coordinates 09:56:45.7
+28:49:35), situated at the periphery of this void at a distance of
D$\sim$6.5 Mpc.
We also present the results of $V,R$, and H$\alpha$ imaging, indicate its
very blue colours, and discuss DDO~68 properties that make it a probable
candidate for a young galaxy.

\begin{figure*}
   \vspace*{1cm}
   \centering
 \includegraphics[angle=0,width=7.5cm, clip=]{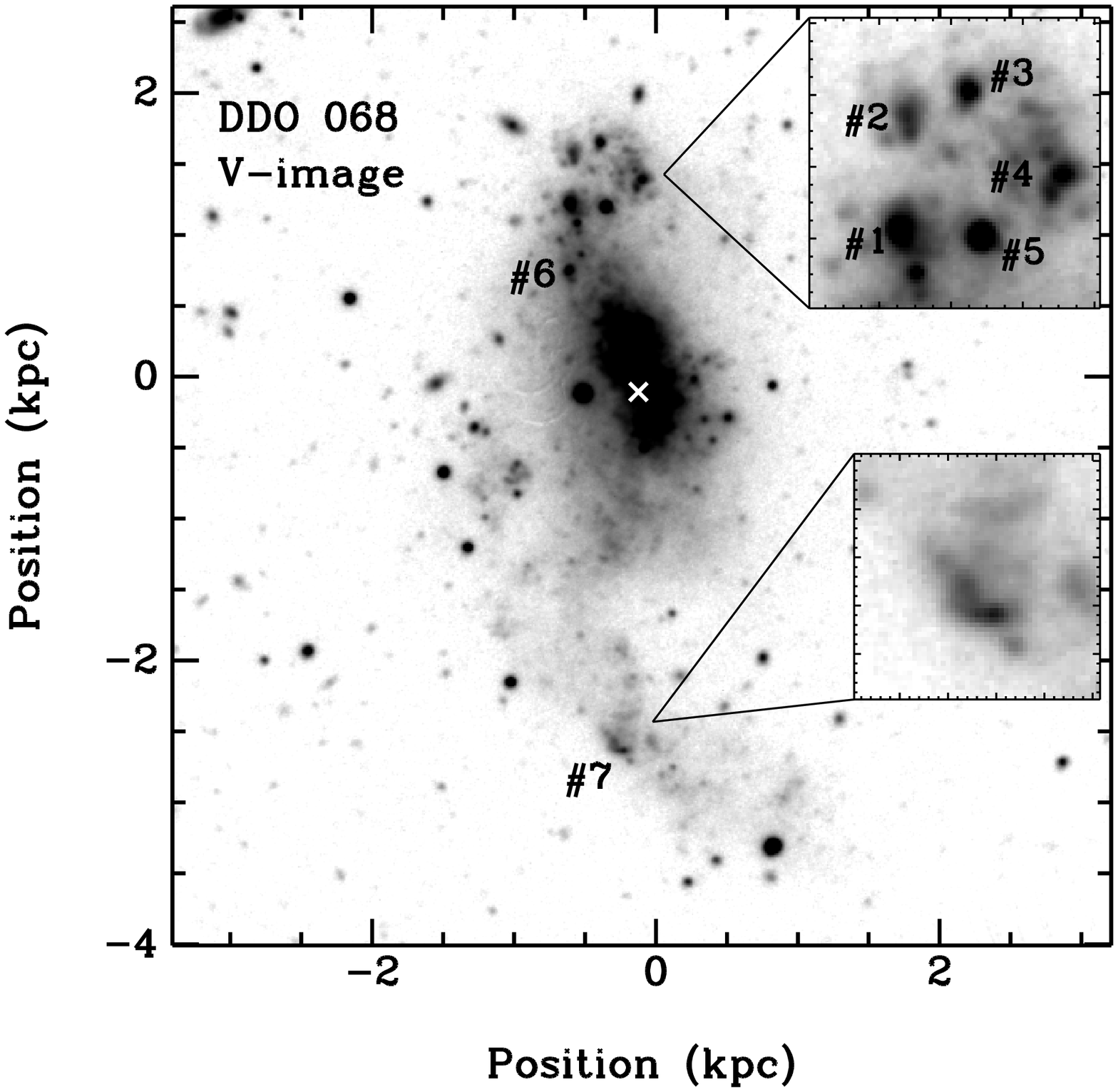}
 \includegraphics[angle=0,width=7.5cm, clip=]{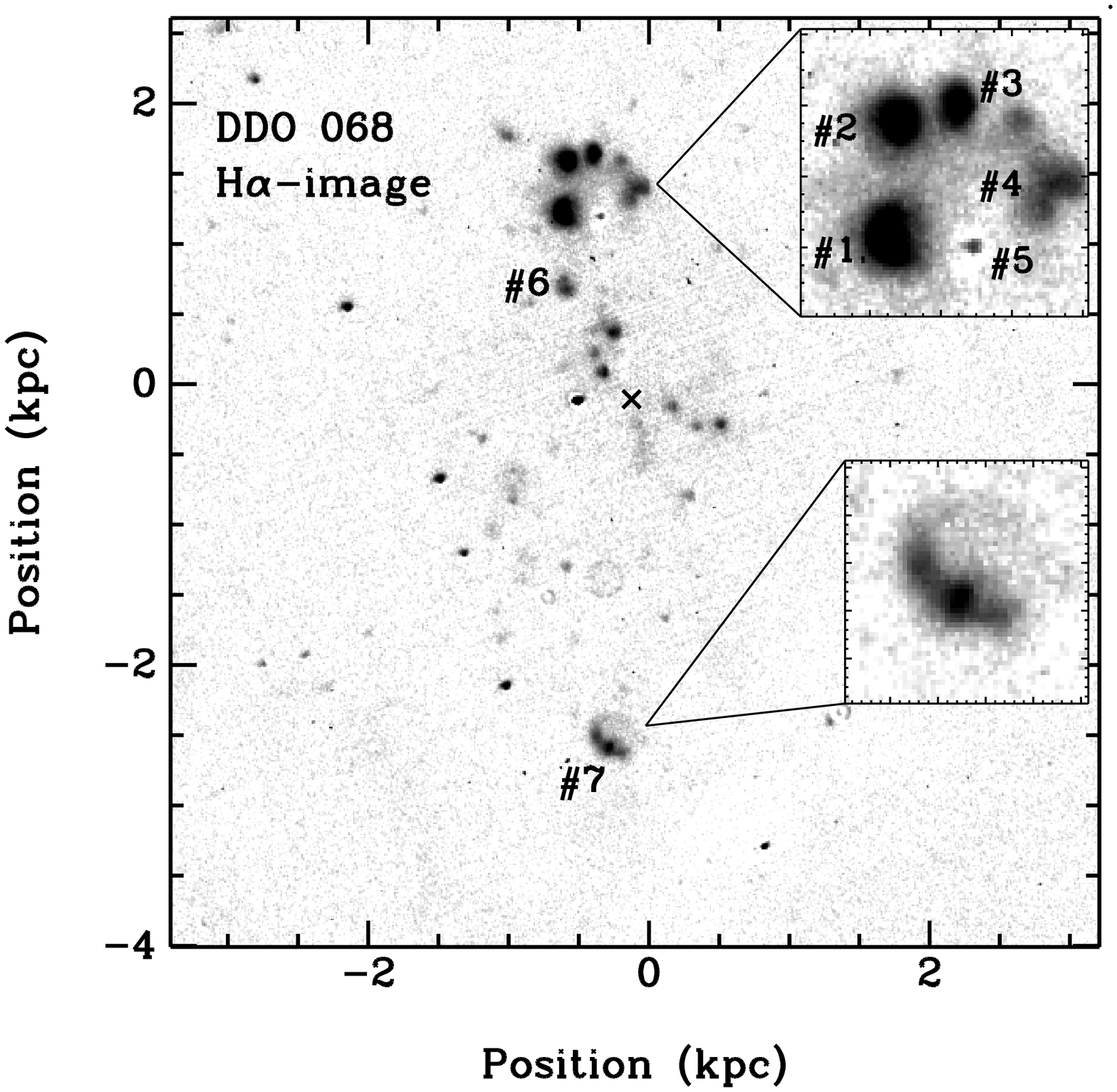}
   \caption{The $V$-band and net H$\alpha$ images of DDO~68 showing its
   overall morphology and the regions of the currently intense star formation.
  The numbers mark the bright knots whose spectra are
   discussed in the paper. The cross marks the central position for building
   the surface brightness profiles in Fig. \ref{ProfVR}. The scale is 31.5 pc
   per 1\arcsec. The insets show blow-ups of the ring-like features.}
	 \label{image_V}
 \end{figure*}

\section{Observations and reduction}
\label{Obs}

\begin{table*}
\begin{center}
\caption{Journal of the 6\,m telescope observations of DDO~68}
\label{Tab1}
\begin{tabular}{lrccccccc} \\ \hline \hline
\MC{1}{c}{ Date }       &
\MC{1}{c}{ Exposure }   &
\MC{1}{c}{ Wavelen.Range [\AA] } &
\MC{1}{c}{ Dispersion } &
\MC{1}{c}{ Seeing }     &
\MC{1}{c}{ Airmass }     &
\MC{1}{c}{ PA }          \\

\MC{1}{c}{ }       &
\MC{1}{c}{ time [s] }    &
\MC{1}{c}{ or Filter } &
\MC{1}{c}{ [\AA/pixel] } &
\MC{1}{c}{ [arcsec] }    &
\MC{1}{c}{          }    &
\MC{1}{c}{ [degree] }     \\

\MC{1}{c}{ (1) } &
\MC{1}{c}{ (2) } &
\MC{1}{c}{ (3) } &
\MC{1}{c}{ (4) } &
\MC{1}{c}{ (5) } &
\MC{1}{c}{ (6) } &
\MC{1}{c}{ (7) } &  \\
\hline
\\[-0.3cm]
 09.11.2004  & 2x900  & $ 3500-7500$  & 2.0 &  0.8 & 1.15 & ~90 \\
 09.11.2004  & 1x900  & $ 3500-7500$  & 2.0 &  0.8 & 1.08 & 123 \\
 08.01.2005  & 4x900  & $ 3500-7500$  & 2.0 &  1.7 & 1.07 & ~~0 \\
 13.01.2005  & 3x900  & $ 3500-7500$  & 2.0 &  1.4 & 1.09 & 123 \\
 12.01.2005  & 3x600  & $V$           & --- &  1.8 & 1.07 & --- \\
 12.01.2005  & 3x600  & $R$           & --- &  1.7 & 1.20 & --- \\
 12.01.2005  & 1x900  & $SED665$      & --- &  1.7 & 1.27 & --- \\
 12.01.2005  & 1x900  & $SED607$      & --- &  1.7 & 1.33 & --- \\
\hline \hline \\[-0.2cm]
\end{tabular}
\end{center}
\end{table*}

All observations were conducted with the SCORPIO multi-mode instrument
(Afanasiev \& Moiseev \cite{SCORPIO})
installed in the prime focus of the SAO 6\,m telescope (BTA), during 2 runs
-- in November 9--11, 2004 and January 7--13, 2005.
For the long-slit spectral observations the grism VPH550g was used with the
2K$\times$2K CCD detector EEV~42-40, with the exposed region of
2048$\times$600 px. This gave the range 3500--7500~\AA\
with $\sim$2.0 \AA~pixel$^{-1}$ and FWHM $\sim$12~\AA\ along the dispersion.
The scale along the slit was 0\farcs18 pixel$^{-1}$ and  total extent of
$\sim$2\arcmin.
The spectra were obtained for several slit positions, crossing the bright
knots, marked by their numbers on the
$V$-band image of the galaxy obtained also in this program
(Fig. \ref{image_V}). Exposure times varied between 15 to 60 min. When
longer than 15 min, they were broken into subexposures of 15-min duration.
The objects spectra were complemented before or after by the reference
spectra of He--Ne--Ar lamp for the wavelength calibration.
Bias and flat-field images were also acquired to perform the standard
reduction of 2D spectra. Spectral standard star Feige~34
(Bohlin \cite{Bohlin96}) was observed during the night for the flux
calibration.

The primary task of these observations was to get high \mbox{S-to-N} spectra
of several \ion{H}{ii} regions in order to detect
the faint line [\ion{O}{iii}]$\lambda$4363 and to directly measure their
electron temperatures T$_{\rm e}$ and oxygen abundances. Seeing was
$\sim$0\farcs8 in the November 2004 run and $\sim$1\farcs4--1\farcs7 in
the January 2005 run.
We also used SCORPIO in the imaging mode (2048$\times$2048 px,
binned 2$\times$2, the field of view of
$\sim$6\arcmin$\times$6\arcmin) to study DDO~68 morphology,
structure and colours. The standard Johnson-Cousins $V,R_{c}$   
and the middle-band filters for H$\alpha$-line images
SED665 (central wavelength = 6622~\AA, FWHM=191~\AA) and SED607
(central wavelength = 6063~\AA, FWHM=167~\AA)  were used.
On the night of January 12, 2005,
we obtained  30-minute exposures of the galaxy in $V$ and $R$ bands
(all broken into 10-min subexposures), and
15-min exposures in both SED665 and SED607 filters. The night was
photometric, with the seeing between 1\farcs6 and 1\farcs8.
Bias and flat-field images were acquired to perform the standard reduction.
For the broad-band calibration, we observed the standard stars during the
night in the field of
QSO FBQ 0951+263 (Nakos et al. \cite{Nakos03}), while for the
calibration of H$\alpha$ images we observed the spectrophotometric standard
Feige~34.
The coefficients of the resulting photometric system have the overall
uncertainties of 0\fm046 in $V$ and  0\fm056 in $R$.
The standard pipeline with the use of IRAF\footnote{IRAF: the Image Reduction
and Analysis Facility is
distributed by the National Optical Astronomy Observatory, which is
operated by the Association of Universities for Research in Astronomy
(AURA) under cooperative agreement with the National Science
Foundation (NSF).}
and MIDAS\footnote{MIDAS is an acronym for the European Southern
Observatory package -- Munich Image Data Analysis System. }
was applied for the reduction of both long-slit
spectra and images, which included the next steps.

Cosmic ray hits were removed from all spectra and direct images in MIDAS.
Using IRAF packages from CCDRED for spectra and images, we subtracted bias,
and performed flat-field correction.
After that, spectra were wavelength-calibrated. 
Night sky background was subtracted from all spectra.
Then, using the data on spectrophotometry standard stars,
all spectra were transformed to absolute fluxes.
One-dimensional spectra were extracted by summing up, without weighting,
various numbers of rows along the slit depending on exact region of
interest.
The next step of the reduction of direct images consisted of shifting all V,
R, SED607, and SED665 images into the common world coordinate system.
Finally, all subexposures were combined into one image by summation.

To obtain the integrated photometry, the $V$ and $R$ surface brightness
profiles (SBPs) and the fitting of SBPs, the method and the programs
described in detail in Kniazev et al. (\cite{LSBs}) were used:
(1) $V$ and $R$ images were combined and filtered with a smooth-and-clip
filter  (Shergin, Kniazev \& Lipovetsky \cite{Sh_Kn_Li_96});
(2) the galaxy was detected above the 3$\sigma$ noise level on the combined
and filtered image and the mask-frame was created to show the location of the
galaxy for the subsequent steps of the analysis;
(3) all background galaxies and stars were additionally masked;
(4) using the mask-frame, the integrated photometry was calculated, the
SBPs were created for each filter with the circular apertures with the step
of 2\arcsec,  and the errors were calculated;
(5) fitting of SBPs by the exponential disc was performed
with weights  $w_k = \sigma_k^{-1}$, where $\sigma_k$ is the error of the
flux calculated within the circular aperture for each SB level
in the previous module.

In the three \ion{H}{ii} regions (Knots 1, 2, and 3 in the Northern ring/oval)
where the line
[\ion{O}{iii}]$\lambda$4363 was seen, we extracted only the regions
along the slit where this line was well above the noise (from 1\farcs8 to
3\farcs8 for various knots). All emission lines were measured by applying the
MIDAS programs
described in detail in a recent paper by Kniazev et al. (\cite{SHOC}).
Briefly, they draw continuum, perform robust noise estimation, fit
separate lines by a single Gaussian superimposed on the continuum-subtracted
spectrum and integrate the line flux. The emission lines, blended in pairs or
triplets, were fitted simultaneously as blend of two or more Gaussian
features. The quoted errors
of singular line intensities include the following components. The first is
related  to the Poisson statistics of the line photon flux. The second
component is the
error resulting from the creation of the underlying continuum, which gives the
main contribution to the errors of faint lines. For the fluxes of the lines
in blends, an additional error appears related to the goodness of fit. Last,
the term related to the uncertainty of the spectral sensitivity curve gives
an additional error to the relative line intensities. This term is 5\%
for the observations presented, hence, it gives the main contribution
to errors of the relative intensities of strong lines.  All these
components are summed squared, and the total errors
have been propagated to calculate the errors of all derived parameters.

\begin{figure*}
   \vspace*{1cm}
   \centering
 \includegraphics[angle=0,width=14cm, clip=]{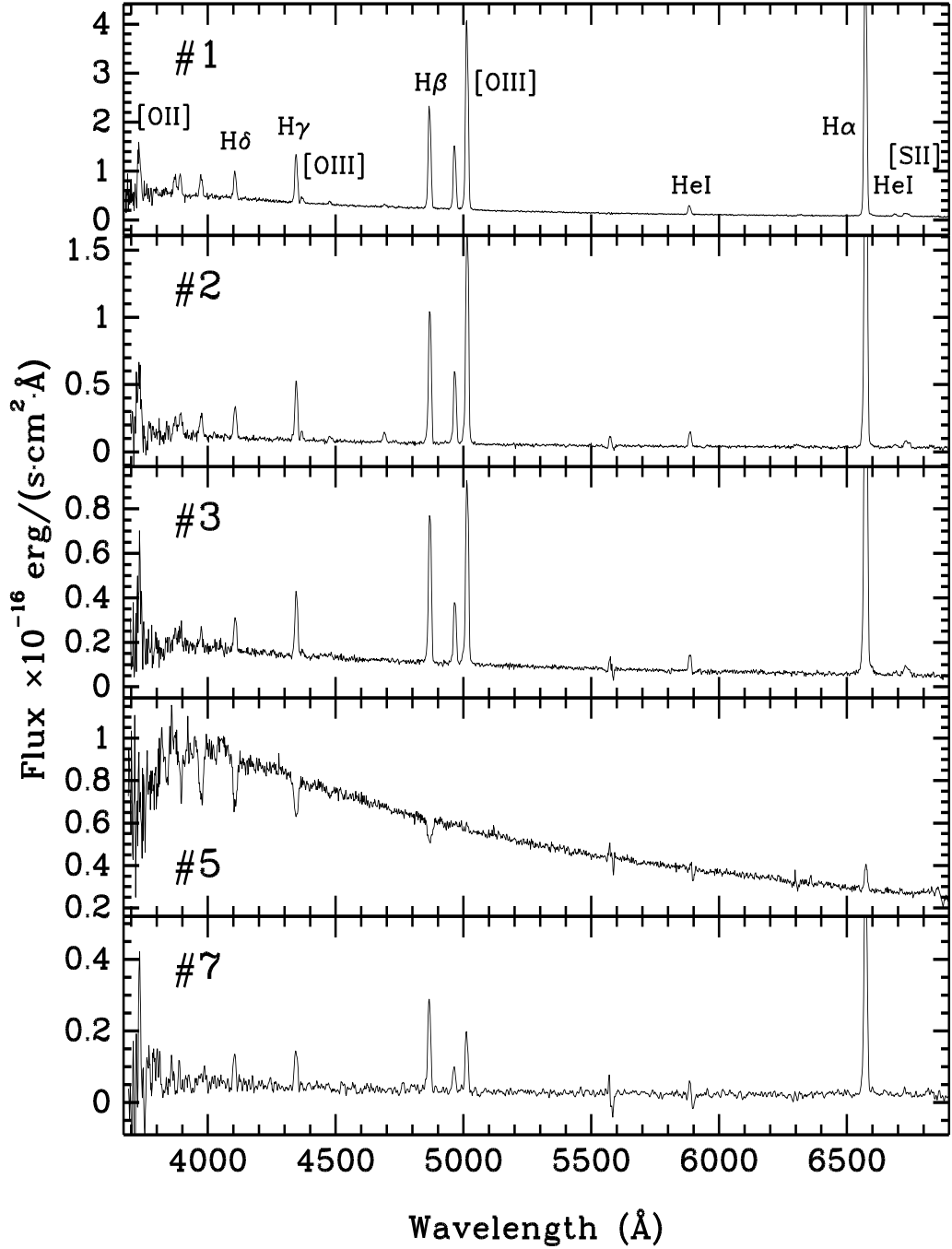}
   \caption{The spectra of 5 knots in DDO~68 (see nomenclature in
   Fig. \ref{image_V}). Main lines are marked on the spectrum of Knot 1.
   The spectrum of Knot 7 (with no reliable O/H) is presented to show
   the general similarity to those of other knots that also indicate
   low O/H.
       }
	 \label{fig:spectra}
 \end{figure*}

\begin{table*}[hbtp]
\centering{
\caption{Line intensities in Knot 1 of DDO~68}
\label{t:Intens1}
\begin{tabular}{lcccccc} \hline  \hline
\rule{0pt}{10pt}
& \MC{2}{c}{PA=90\degr,  $\theta$=0\farcs8}   && \MC{2}{c}{PA=0\degr, $\theta$=1\farcs6} \\ \cline{2-3} \cline{5-6}
\rule{0pt}{10pt}
$\lambda_{0}$(\AA) Ion                    &
$F$($\lambda$)/$F$(H$\beta$)&$I$($\lambda$)/$I$(H$\beta$) &&
$F$($\lambda$)/$F$(H$\beta$)&$I$($\lambda$)/$I$(H$\beta$) \\ \hline

3727\ [O\ {\sc ii}]\            & 0.196$\pm$0.009 & 0.190$\pm$0.011  && 0.552$\pm$0.028 & 0.545$\pm$0.030   \\
3869\ [Ne\ {\sc iii}]\          & ...             & ...              && 0.184$\pm$0.016 & 0.182$\pm$0.017   \\
3967\ [Ne\ {\sc iii}]\ +\ H7\   & 0.188$\pm$0.011 & 0.253$\pm$0.019  && 0.215$\pm$0.011 & 0.235$\pm$0.019    \\
4101\ H$\delta$\                & 0.201$\pm$0.011 & 0.266$\pm$0.019  && 0.242$\pm$0.013 & 0.260$\pm$0.019   \\
4340\ H$\gamma$\                & 0.423$\pm$0.023 & 0.473$\pm$0.029  && 0.466$\pm$0.024 & 0.477$\pm$0.027   \\
4363\ [O\ {\sc iii}]\           & 0.047$\pm$0.07 & 0.046$\pm$0.007   && 0.058$\pm$0.007 & 0.057$\pm$0.007   \\
4686\ He\ {\sc ii}\             & 0.022$\pm$0.003 & 0.022$\pm$0.003  && 0.033$\pm$0.004 & 0.033$\pm$0.004   \\
4861\ H$\beta$\                 & 1.000$\pm$0.005 & 1.000$\pm$0.006  && 1.000$\pm$0.004 & 1.000$\pm$0.005   \\
4959\ [O\ {\sc iii}]\           & 0.646$\pm$0.033 & 0.616$\pm$0.033  && 0.624$\pm$0.032 & 0.616$\pm$0.032   \\
5007\ [O\ {\sc iii}]\           & 1.933$\pm$0.097 & 1.838$\pm$0.096  && 1.899$\pm$0.095 & 1.876$\pm$0.095   \\
5876\ He\ {\sc i}\              & 0.102$\pm$0.007 & 0.093$\pm$0.006  && 0.079$\pm$0.006 & 0.078$\pm$0.006   \\
6300\ [O\ {\sc i}]\             & 0.021$\pm$0.003 & 0.019$\pm$0.003  && ...             & ...               \\
6312\ [S\ {\sc iii}]\           & 0.010$\pm$0.003 & 0.009$\pm$0.003  && ...             & ...               \\
6563\ H$\alpha$\                & 3.102$\pm$0.159 & 2.772$\pm$0.161  && 2.740$\pm$0.139 & 2.711$\pm$0.151   \\
6584\ [N\ {\sc ii}]\            & ...             & ...              && 0.017$\pm$0.034 & 0.016$\pm$0.034   \\
6678\ He\ {\sc i}\              & 0.024$\pm$0.004 & 0.021$\pm$0.003  && 0.024$\pm$0.004 & 0.024$\pm$0.004   \\
6717\ [S\ {\sc ii}]\            & 0.035$\pm$0.006 & 0.031$\pm$0.005  && 0.034$\pm$0.005 & 0.033$\pm$0.005   \\
6731\ [S\ {\sc ii}]\            & 0.021$\pm$0.006 & 0.019$\pm$0.005  && 0.018$\pm$0.004 & 0.018$\pm$0.004   \\
7136\ [Ar\ {\sc iii}]\          & 0.026$\pm$0.004 & 0.023$\pm$0.004  && 0.019$\pm$0.003 & 0.019$\pm$0.008    \\
C(H$\beta$)\ dex          & \MC {2}{c}{0.10$\pm$0.07} && \MC {2}{c}{0.00$\pm$0.07}  \\
EW(abs)\ \AA\             & \MC {2}{c}{5.05$\pm$0.82} && \MC {2}{c}{1.35$\pm$0.75}  \\
$F$(H$\beta$)$^a$\        & \MC {2}{c}{24.8$\pm$1.3}     && \MC {2}{c}{27.2$\pm$1.4}      \\
EW(H$\beta$)\ \AA\        & \MC {2}{c}{ 117$\pm$0.6}   && \MC {2}{c}{ 110$\pm$0.5}    \\
Rad. vel.\ \kms\          & \MC {2}{c}{ 450$\pm$48}   && \MC {2}{c}{ 378$\pm$33}    \\
\hline  \hline
\MC{5}{l}{$^a$ in units of 10$^{-16}$ ergs\ s$^{-1}$cm$^{-2}$.}\\
\end{tabular}
 }
\end{table*}

\begin{table*}[hbtp]
\centering{
\caption{Line intensities in Knots 2, 3 and 7 of DDO~68}
\label{t:Intens2}
\begin{tabular}{lcccccccc} \hline  \hline
\rule{0pt}{10pt}
& \MC{2}{c}{Knot 2, PA=--57\degr, $\theta$=0\farcs8}   && \MC{2}{c}{Knot 2, PA=0\degr, $\theta$=1\farcs6 } && \MC{2}{c}{Knot 2, PA=--57\degr, $\theta$=1\farcs4}  \\ \cline{2-3} \cline{5-6} \cline{8-9}
\rule{0pt}{10pt}
$\lambda_{0}$(\AA) Ion                    &
$F$($\lambda$)/$F$(H$\beta$)&$I$($\lambda$)/$I$(H$\beta$) &&
$F$($\lambda$)/$F$(H$\beta$)&$I$($\lambda$)/$I$(H$\beta$) &&
$F$($\lambda$)/$F$(H$\beta$)&$I$($\lambda$)/$I$(H$\beta$) \\ \hline

3727\ [O\ {\sc ii}]\           & 0.464$\pm$0.025 & 0.467$\pm$0.027 && 0.451$\pm$0.023 & 0.451$\pm$0.025  && 0.698$\pm$0.035 & 0.798$\pm$0.045    \\
3967\ [Ne\ {\sc iii}]\ +\ H7\  & ...             & ...             && 0.221$\pm$0.013 & 0.221$\pm$0.019  && 0.176$\pm$0.011 & 0.227$\pm$0.021    \\
4101\ H$\delta$\               & 0.247$\pm$0.022 & 0.263$\pm$0.023 && 0.264$\pm$0.015 & 0.264$\pm$0.020  && 0.230$\pm$0.013 & 0.280$\pm$0.022    \\
4340\ H$\gamma$\               & 0.470$\pm$0.033 & 0.474$\pm$0.046 && 0.486$\pm$0.026 & 0.486$\pm$0.028  && 0.424$\pm$0.024 & 0.469$\pm$0.029    \\
4363\ [O\ {\sc iii}]\          & 0.049$\pm$0.018 & 0.049$\pm$0.018 && 0.049$\pm$0.008 & 0.049$\pm$0.008  && 0.051$\pm$0.008 & 0.053$\pm$0.009    \\
4686\ He\ {\sc ii}\            & 0.121$\pm$0.015 & 0.120$\pm$0.015 && 0.096$\pm$0.007 & 0.096$\pm$0.007  && 0.082$\pm$0.008 & 0.082$\pm$0.009    \\
4861\ H$\beta$\                & 1.000$\pm$0.017 & 1.000$\pm$0.022 && 1.000$\pm$0.007 & 1.000$\pm$0.007  && 1.000$\pm$0.007 & 1.000$\pm$0.008    \\
4959\ [O\ {\sc iii}]\          & 0.517$\pm$0.031 & 0.515$\pm$0.031 && 0.524$\pm$0.027 & 0.524$\pm$0.027  && 0.537$\pm$0.028 & 0.522$\pm$0.028    \\
5007\ [O\ {\sc iii}]\          & 1.538$\pm$0.078 & 1.532$\pm$0.078 && 1.548$\pm$0.077 & 1.548$\pm$0.078  && 1.620$\pm$0.081 & 1.564$\pm$0.080    \\
5876\ He\ {\sc i}\             & 0.087$\pm$0.012 & 0.086$\pm$0.012 && 0.061$\pm$0.005 & 0.061$\pm$0.005  && 0.087$\pm$0.007 & 0.077$\pm$0.006    \\
6300\ [O\ {\sc i}]\            & ...             & ...             && ...             & ...              && 0.020$\pm$0.010 & 0.017$\pm$0.009    \\
6312\ [S\ {\sc iii}]\          & ...             & ...             && ...             & ...              && 0.015$\pm$0.012 & 0.013$\pm$0.011    \\
6563\ H$\alpha$\               & 2.777$\pm$0.140 & 2.740$\pm$0.151 && 2.651$\pm$0.133 & 2.651$\pm$0.144  && 3.286$\pm$0.167 & 2.744$\pm$0.155    \\
6678\ He\ {\sc i}\             & 0.034$\pm$0.015 & 0.030$\pm$0.015 && 0.012$\pm$0.004 & 0.012$\pm$0.004  && 0.028$\pm$0.007 & 0.023$\pm$0.006    \\
6717\ [S\ {\sc ii}]\           & 0.066$\pm$0.017 & 0.065$\pm$0.017 && 0.035$\pm$0.007 & 0.035$\pm$0.007  && 0.056$\pm$0.009 & 0.046$\pm$0.008    \\
6731\ [S\ {\sc ii}]\           & ...             & ...             && 0.022$\pm$0.006 & 0.022$\pm$0.006  && 0.027$\pm$0.007 & 0.022$\pm$0.006    \\
7136\ [Ar\ {\sc iii}]\         & ...             & ...             && 0.028$\pm$0.009 & 0.028$\pm$0.009  && 0.021$\pm$0.007 & 0.017$\pm$0.005    \\
C(H$\beta$)\ dex          & \MC {2}{c}{0.02$\pm$0.07} && \MC {2}{c}{0.00$\pm$0.07} && \MC {2}{c}{0.22$\pm$0.07}  \\
EW(abs)\ \AA\             & \MC {2}{c}{0.75$\pm$6.18} && \MC {2}{c}{0.00$\pm$2.24} && \MC {2}{c}{3.70$\pm$1.55}  \\
$F$(H$\beta$)$^a$\        & \MC {2}{c}{12.1$\pm$0.7}  && \MC {2}{c}{13.7$\pm$0.7}  && \MC {2}{c}{13.6$\pm$0.7}      \\
EW(H$\beta$)\ \AA\        & \MC {2}{c}{ 286$\pm$5}    && \MC {2}{c}{ 299$\pm$2}    && \MC {2}{c}{ 198$\pm$2}     \\
Rad. vel.\ \kms\          & \MC {2}{c}{ 528$\pm$33}   && \MC {2}{c}{ 411$\pm$33}   && \MC {2}{c}{ 447$\pm$45}    \\
\hline  \hline
\rule{0pt}{10pt}
& \MC{2}{c}{Knot 3, PA=--57\degr,  $\theta$=0\farcs8}   && \MC{2}{c}{Knot 3, PA=--57\degr, $\theta$=1\farcs4} && \MC{2}{c}{Knot 7, PA=40\degr, $\theta$=1\farcs2}\\ \cline{2-3} \cline{5-6} \cline{8-9}
\rule{0pt}{10pt}
$\lambda_{0}$(\AA) Ion                    &
$F$($\lambda$)/$F$(H$\beta$)&$I$($\lambda$)/$I$(H$\beta$) &&
$F$($\lambda$)/$F$(H$\beta$)&$I$($\lambda$)/$I$(H$\beta$) &&
$F$($\lambda$)/$F$(H$\beta$)&$I$($\lambda$)/$I$(H$\beta$) \\ \hline

3727\ [O\ {\sc ii}]\   & 0.656$\pm$0.034 & 0.678$\pm$0.040  && 0.656$\pm$0.033 & 0.823$\pm$0.045 && 1.085$\pm$0.082 & 1.194$\pm$0.095   \\
4101\ H$\delta$\       & 0.205$\pm$0.016 & 0.278$\pm$0.028  && 0.211$\pm$0.013 & 0.249$\pm$0.021 && 0.388$\pm$0.089 & 0.413$\pm$0.109   \\
4340\ H$\gamma$\       & 0.421$\pm$0.030 & 0.474$\pm$0.038  && 0.454$\pm$0.027 & 0.503$\pm$0.032 && 0.465$\pm$0.059 & 0.484$\pm$0.094   \\
4363\ [O\ {\sc iii}]\  & 0.061$\pm$0.018 & 0.060$\pm$0.018  && 0.039$\pm$0.010 & 0.042$\pm$0.011 && 0.029$\pm$0.038 & 0.031$\pm$0.040   \\
4861\ H$\beta$\        & 1.000$\pm$0.016 & 1.000$\pm$0.018  && 1.000$\pm$0.009 & 1.000$\pm$0.010 && 1.000$\pm$0.038 & 1.000$\pm$0.042   \\
4959\ [O\ {\sc iii}]\  & 0.438$\pm$0.026 & 0.418$\pm$0.026  && 0.412$\pm$0.022 & 0.403$\pm$0.022 && 0.212$\pm$0.029 & 0.210$\pm$0.029   \\
5007\ [O\ {\sc iii}]\  & 1.329$\pm$0.069 & 1.264$\pm$0.069  && 1.270$\pm$0.064 & 1.235$\pm$0.063 && 0.600$\pm$0.032 & 0.593$\pm$0.032   \\
5876\ He\ {\sc i}\     & 0.094$\pm$0.012 & 0.085$\pm$0.011  && 0.088$\pm$0.008 & 0.075$\pm$0.007 && 0.118$\pm$0.018 & 0.110$\pm$0.017   \\
6548\ [N\ {\sc ii}]\   & ...             & ...              && 0.010$\pm$0.007 & 0.008$\pm$0.005 && 0.006$\pm$0.023 & 0.005$\pm$0.021   \\
6563\ H$\alpha$\       & 3.019$\pm$0.152 & 2.690$\pm$0.154  && 3.535$\pm$0.179 & 2.750$\pm$0.152 && 2.977$\pm$0.062 & 2.683$\pm$0.062   \\
6584\ [N\ {\sc ii}]\   & ...             & ...              && 0.035$\pm$0.020 & 0.027$\pm$0.016 && 0.029$\pm$0.035 & 0.026$\pm$0.032   \\
6678\ He\ {\sc i}\     & 0.026$\pm$0.009 & 0.023$\pm$0.008  && 0.024$\pm$0.006 & 0.019$\pm$0.004 && ...             & ...               \\
6717\ [S\ {\sc ii}]\   & 0.056$\pm$0.018 & 0.050$\pm$0.016  && 0.066$\pm$0.012 & 0.051$\pm$0.009 && 0.056$\pm$0.024 & 0.050$\pm$0.021   \\
6731\ [S\ {\sc ii}]\   & 0.019$\pm$0.015 & 0.017$\pm$0.014  && 0.035$\pm$0.011 & 0.027$\pm$0.009 && ...             & ...               \\
C(H$\beta$)\ dex          & \MC {2}{c}{0.10$\pm$0.07} && \MC {2}{c}{0.32$\pm$0.07} && \MC {2}{c}{0.14$\pm$0.07}  \\
EW(abs)\ \AA\             & \MC {2}{c}{3.20$\pm$0.61} && \MC {2}{c}{0.25$\pm$0.63} && \MC {2}{c}{0.05$\pm$5.71}  \\
$F$(H$\beta$)$^a$\        & \MC {2}{c}{8.5$\pm$0.5}   && \MC {2}{c}{9.1$\pm$0.5}   && \MC {2}{c}{3.4$\pm$0.2}     \\
EW(H$\beta$)\ \AA\        & \MC {2}{c}{ 75$\pm$ 1}    && \MC {2}{c}{ 77$\pm$ 1}    && \MC {2}{c}{ 86$\pm$ 3}   \\
Rad. vel.\ \kms\          & \MC {2}{c}{ 501$\pm$36}   && \MC {2}{c}{ 447$\pm$39}   && \MC {2}{c}{ 309$\pm$93}   \\
\hline  \hline
\MC{5}{l}{$^a$ in units of 10$^{-16}$ ergs\ s$^{-1}$cm$^{-2}$.}\\
\end{tabular}
 }
\end{table*}

\begin{table*}[hbtp]
\centering{
\caption{Abundances in Knots 1, 2 and 3 of DDO~68}
\label{t:Chem1}
\begin{tabular}{lcc|ccc|cc} \hline  \hline
\rule{0pt}{10pt}
					   &        \MC{2}{c}{Knot 1}            &          \MC{3}{c}{Knot 2}                                 &          \MC{2}{c}{Knot 3} \\ \cline{2-3} \cline{4-6} \cline{7-8}
\rule{0pt}{10pt}
Value                                      & PA=--90\degr    & PA=0\degr         & PA=--57\degr       & PA=0\degr         & PA=--57\degr      & PA=--57\degr      & PA=--57\degr      \\
					   &$\theta$=0\farcs8&$\theta$=1\farcs6  & $\theta$=0\farcs8  & $\theta$=1\farcs6 & $\theta$=1\farcs4 & $\theta$=0\farcs8 & $\theta$=1\farcs4 \\ \hline
$T_{\rm e}$(OIII)(10$^{3}$~K)\             & 17.01$\pm$1.32  & 18.89$\pm$1.33    & 19.33$\pm$4.27  & 19.24$\pm$1.90 & 20.13$\pm$2.07          & 25.30$\pm$5.92  & 20.28$\pm$3.28   \\
$T_{\rm e}$(OII)(10$^{3}$~K)\              & 14.64$\pm$1.08  & 15.34$\pm$1.02    & 15.48$\pm$3.18  & 15.45$\pm$1.42 & 15.72$\pm$1.48          & 16.73$\pm$2.61  & 15.77$\pm$2.33   \\
$T_{\rm e}$(SIII)(10$^{3}$~K)\             & 15.82$\pm$1.1   & 17.38$\pm$1.11    & 17.73$\pm$3.54  & 17.66$\pm$1.58 & 18.41$\pm$1.72          & 22.70$\pm$4.92  & 18.53$\pm$2.72   \\
& & & & & & &\\
O$^{+}$/H$^{+}$($\times$10$^{-5}$)\        & 0.177$\pm$0.036  & 0.442$\pm$0.078  & 0.369$\pm$0.193 & 0.358$\pm$0.086& 0.602$\pm$0.145         & 0.431$\pm$0.159 & 0.616$\pm$0.229  \\
O$^{++}$/H$^{+}$($\times$10$^{-5}$)\       & 1.489$\pm$0.287  & 1.205$\pm$0.203  & 0.944$\pm$0.479 & 0.965$\pm$0.222& 0.886$\pm$0.208         & 0.470$\pm$0.230 & 0.686$\pm$0.251  \\
O$^{+++}$/H$^{+}$($\times$10$^{-5}$)\      & 0.052$\pm$0.015  & 0.068$\pm$0.016  & 0.192$\pm$0.074 & 0.309$\pm$0.110& 0.159$\pm$0.053         & ...             & ...              \\
O/H($\times$10$^{-5}$)\                    & 1.716$\pm$0.289  & 1.715$\pm$0.218  & 1.505$\pm$0.521 & 1.632$\pm$0.262& 1.648$\pm$0.260         & 0.901$\pm$0.280 & 1.302$\pm$0.339  \\
12+log(O/H)\                               & ~7.23$\pm$0.07~  & ~7.23$\pm$0.06~  & ~7.18$\pm$0.16~ & ~7.21$\pm$0.07~& ~7.22$\pm$0.07~         & ~6.95$\pm$0.14~ & ~7.11$\pm$0.11~  \\
\hline   \hline
\end{tabular}
 }
\end{table*}

\section{Results}
\label{results}

\subsection{Line intensities and element abundances}
\label{abun}

The relative intensities of all emission lines used for abundance
determination in the 3 discussed \ion{H}{ii} regions (Knots 1, 2, and 3),
as well as the derived CH$\beta$, EWs of Balmer absorption lines, measured
flux in H$\beta$ emission line, and the measured heliocentric radial
velocities of each knot, are given in Tables \ref{t:Intens1} and
\ref{t:Intens2}. The position angle (PA) of
the slit and the seeing $\theta$ during observation are also given for all
spectra.
The best 1D spectra for each of these regions, as well as the spectra of
Knots 5 and 7, are shown in Fig.~\ref{fig:spectra}.
For each of Knots 1, 2, and 3, we got repeated observations and can compare
individual data. For Knot 1 the relative line intensities are very consistent
in the two spectra for the most of lines except [\ion{O}{ii}]$\lambda$3727 and
H$\alpha$. This presumably is due to the
different slit position on the asymmetric \ion{H}{ii} region and also due to
a factor of two different seeings, resulting in some contribution from the
outer parts of this knot.
For Knot 2 the first two spectra are very consistent with each other, while
the 3-d indicates some additional `reddening'. The similar situation is for
the 2nd spectrum of Knot 3 (PA=--57$\degr$, $\theta$=1\farcs4), extracted
from the same 2D spectrum. This seems to indicate some additional error in the
spectral sensitivity curve for this night. However, the effect of this on the
resulting abundances is small.

Extinction in the observed \ion{H}{ii} regions is low: C(H$\beta$)
$\lesssim$0.1.
A larger value of C(H$\beta$), derived for one spectrum of Knots 2 and 3
($\sim$0.2--0.3), is probably due to the worse quality of the spectral
sensitivity curve for this night. This low C(H$\beta$) is quite
consistent with the extinction data for most very metal-poor galaxies.

Chemical abundances and physical parameters are determined in the frame of
the classical  two-zone model of \ion{H}{ii} region (Stasi\'nska
\cite{Stas90}), as described, e.g.,
in our recent papers by Pustilnik et al. (\cite{HS0837}) and by Kniazev et
al. (\cite{SHOC}), which in turn follow the method described by Izotov
et al. (\cite{ITL97}).
The derived individual abundances of oxygen for Knots 1, 2, and 3,
as well as their T$_{\rm e}$, are given in Table~\ref{t:Chem1}.
Since the [\ion{S}{ii}] $\lambda$6717/$\lambda$6731 line ratio
is higher than 1.4 for all spectra, the density N$_{e}$ = 10 cm$^{-3}$
was accepted in all our calculations (Aller \cite{Aller}).

Since the line fluxes from these knots are rather small, the S-to-N in the
principal line [\ion{O}{iii}]$\lambda$4363 was at most $\sim$8. Therefore
the majority of resulting abundances for individual observations are of
medium  accuracy. However, they all are consistent each to the other
within their cited uncertainties. Thus, we average them and accept the mean
values for each of the three regions:
12+$\log$(O/H)=7.23$\pm$0.05, 7.21$\pm$0.05, and
7.03$\pm$0.08, for Knots 1, 2, and 3,  respectively.
These mean values for different knots are consistent each to the other within
their uncertainties, so we take their mean as a measure of
the heavy element abundances of DDO~68 (at least, in this northern `ring').
If all seven measurements of O/H in these 3 knots are considered as the
independent representatives
of a unique value for all three knots, their weighted mean corresponds to
12+$\log$(O/H)=7.17$\pm$0.03.
The 1st measurement for Knot 3, with 12+$\log$(O/H)=6.95$\pm$0.14, looks,
however, a bit outlying. If we exclude this
from the general mean, then the resulting mean value of the remaining six
measurements (which probably is more conservative) corresponds to
12+$\log$(O/H)=7.21$\pm$0.03.
We take this O/H as a characteristic value of DDO 68 in further discussion.

For Knot 1 the estimates of Ne, S, and Ar abundances were also derived.
They correspond to $\log$(Ne/O)=--0.70$\pm$0.10, $\log$(S/O)=--1.21$\pm$0.16,
and $\log$(Ar/O)=--2.20$\pm$0.07.
Within rather large uncertainties, they are consistent with the abundance
ratios for the most metal-deficient blue compact galaxies (BCGs) presented
by Izotov \& Thuan (\cite{IT99}).

Having the EWs of emission line H$\beta$, we can estimate the ages of
starbursts in these three \ion{H}{ii} regions apparently placed along a more
or less regular oval.
The average values of EW(H$\beta$) for Knots 1--3
are given in Table \ref{t:ages}.
For Knots 4, 6, and 7, we got spectra of relatively low S-to-N, so we do not
discuss them in detail. However, their EWs(H$\beta$) are well measured
and can also be used to date the starbursts at their respective locations.

We use  the latest version (v5) of
the models from Starburst99 (Leitherer et al. \cite{S99},
Vazquez \& Leitherer \cite{S99v5})     with the
lowest metallicity given their ($z$=0.001) to estimate ages of instantaneous
starbursts from the observed EWs(H$\beta$). For the Salpeter IMF with
M$_{\rm low}$ and M$_{\rm up}$ of  0.1~M\sunn\ and 100~M\sunn,
respectively,
the derived ages of all emission knots are also given in Table \ref{t:ages}.
They range from 3 to 7 Myr.
An independent  check of the ages of all knots can be performed using
the EW(H$\alpha$) from the obtained H$\alpha$ images.
Both age estimates are very consistent to each other.

Knot 5 (see Fig. \ref{image_V}) appeared to be a compact young stellar
cluster with well-seen Balmer series in absorption, besides H$\alpha$, on
which some emission is superimposed. The radial velocity of this cluster
derived on these lines (V$_{\rm hel}$=396$\pm$90~\kms)
is quite consistent with those for the \ion{H}{ii} regions
in the `ring' and with the systemic velocity of DDO 68.
The measured EWs of absorption lines H$\beta$, H$\gamma$, H$\delta$  are in
the range of 4.4--5.8~\AA\ (see Table \ref{t:absorption}).
The value of EW(emis. H$\alpha$) after correction for the underlying
absorption is 9.5$\pm$0.5~\AA. The EW of absorption H$\beta$, in turn, after
the correction for emission component, is 5.8$\pm$0.5~\AA.
EWs of H$_{8}$ and H$_{9}$ are measured with large uncertainties and are not
taken for comparison.
We compared all EWs of Balmer absorptions with the respective model values,
presented by Gonzalez Delgado et al. (\cite{Rosa}) for the cluster with
metallicity $z=0.001$ (nearest to the observed one for DDO~68), assuming the
instantaneous starburst with the Salpeter IMF with M$_{\rm low}$ and
M$_{\rm up}$ of 1 and 80 M\sunn. The EWs of emission H$\alpha$ and the
extrapolated one of H$\beta$ were compared to predictions of the latest
version of Starburst99 (Vazquez \& Leitherer \cite{S99v5}) for the Salpeter
IMF with M$_{\rm up}$ = 100~M\sunn.
Both absorption and emission line EW values are consistent with the cluster
age of T$_{\rm cluster}$ $\sim$22--23 Myr.

Thus, we conclude that starbursts in 4 sites along the `ring' are synchronized
within the interval of 2 Myr. Only in Knot 5 did the star formation episode
begin
$\sim$18 Myr earlier than in the rest positions. One of the possible options
for producing such a configuration of starbursts is the induced gas collapse
behind the front of shock wave, generated by the previous strong starburst
(e.g., van Dyk et al. \cite{Puche}, Efremov et al. \cite{Efremov02},
Elmegreen et al. \cite{Elmegreen04}). However, the outlying age of
Knot 5 suggests more complicated SF scenario in this region.

\subsection{Results of imaging}
\label{images}

The best quality image of DDO~68 obtained in these observations is the one in
$V$-band.
This is shown in Fig. \ref{image_V} with the surface brightness cuts allowing
the maximal extent of the galaxy to be followed. Also on this image we mark
the knots for which the spectra, discussed in the previous section,
were acquired. The net H$\alpha$ image in Fig. \ref{image_V} displays the
regions of current SF. We emphasise that besides the
Northern `ring' with diameter of $\sim$500 pc and rather bright emission-line
knots along this `ring', there is a similar structure in the Southern part of
the galaxy that is connected with the Southern tail: Knot 7 in Fig.
\ref{image_V}. As seen on the net H$\alpha$ image,
it has the form of an almost closed `circle' with a diameter of $\sim$270 pc.
Its SE sector is significantly
brighter than the rest of this `circle' and consists of 3 adjacent knots.
The brightest of them is quite elongated with the major axis oriented
perpendicular to the arc to which it is connected. As discussed
above, its starburst age is $\sim$4.5 Myr, which implies
that the formation of the `ring'-like structures at both edges
of DDO 68 have been synchronized by some global process.

\begin{figure}[hbtp]
   \centering
\resizebox{\hsize}{!}{\includegraphics[angle=-90,clip=]{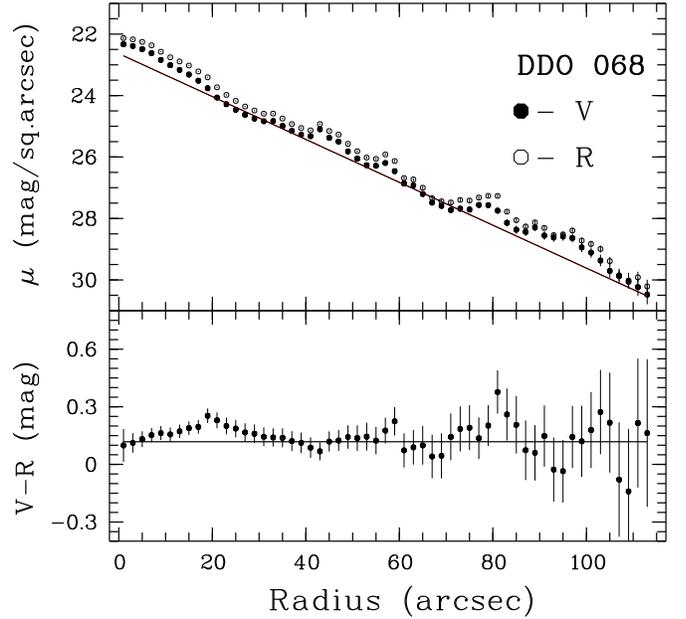}}
   \caption{
{\it Top:} The $V$ and $R$ surface brightness radial profiles (SBP) of
DDO~68 obtained in circular apertures with a step of 2\arcsec.
Shown error bars correspond to 2$\sigma$ uncertainties.
The errors of the photometric system are not included.
The fit by an underlying exponential disk is shown only for $V$-band SBP.
Additional light seen at
$R \sim$40\arcsec\--60\arcsec\ corresponds to Knots No.1--5.
Additional light for $R > $70\arcsec\
corresponds to Knot No.7 and the southern tail.
{\it Bottom:} Respective radial profile of $(V-R)^{0}$ colour with the
correction of $-$0\fm09 for the zero-point shift (see Sect. \ref{correct})
and for the Galaxy extinction.
Error bars correspond to 2$\sigma$ uncertainties without the
errors of the photometric system.
The $(V-R)^{0}$ colour of the exponential disk of 0\fm12 is shown by the
solid line. Its rms uncertainty is $\sigma$=0\fm04.
   }
	 \label{ProfVR}
 \end{figure}

The main issue to address based on our $V,R,$ and H$\alpha$
images is the age of the oldest visible stellar population.
We built surface brightness profiles (SBPs) for $V,R$
images applying aperture photometry in circular apertures centred at
the point near the geometrical centre of the main DDO 68 body (with J2000
coordinates of 09:56:45.8 +28:49:27).
They are shown in Fig.~\ref{ProfVR}.

\begin{table}[hbtp]
\centering{
\caption{Starburst ages in DDO~68}
\label{t:ages}
\begin{tabular}{lcc} \hline  \hline
\rule{0pt}{10pt}
Region                                     & EW(H$\beta$)  & Age (Myr)         \\ \hline
Knot 1                                     & 117$\pm$5     & 4.0 \\
Knot 2                                     & 250$\pm$30    & 3.0 \\
Knot 3                                     &  76$\pm$3     & 5.0 \\
Knot 4                                     & 206$\pm$32    & 3.5 \\
Knot 6                                     &  47$\pm$3     & 7.0 \\
Knot 7                                     &  87$\pm$4     & 4.5 \\
\hline   \hline
\end{tabular}
 }
\end{table}

\begin{table}[hbtp]
\centering{
\caption{Balmer absorptions in Knot 5}
\label{t:absorption}
\begin{tabular}{lr} \hline  \hline
\rule{0pt}{10pt}
$\lambda_{0}$, name       & EW(\AA)         \\ \hline
4102~H$\delta$            & 5.8$\pm$0.5     \\
4340~H$\gamma$            & 4.4$\pm$0.4     \\
4861~H$\beta$             & 4.4$\pm$0.4     \\
6563~H$\alpha$            &$-$4.6$\pm$0.1   \\
\hline   \hline
\MC{2}{l}{Minus in EW(H$\alpha$) means emission} \\
\end{tabular}
 }
\end{table}

The nebular emission in DDO~68 is not as strong and widespread
as in SBS 0335--052 and I~Zw~18. However, it somewhat affects the observed
colours on the significant part of the colour profile in Fig. \ref{ProfVR}.
The contribution of various features superimposed on the smooth distribution
of the assumed underlying light is seen clearly on both SBPs (top
panel) and the colour radial profile (bottom panel).
The strongest features are seen clearly in the H$\alpha$ image (Fig.
\ref{image_V}). The nebular emission shows the most contrast in the two
ring-like regions at the N and S edges of the galaxy. They appear on both
the SBP and
the colour radial distribution as clear peaks at $R \sim$40, 55, 80, and
100\arcsec.
Their contribution within the main body ($R \leq R_{\rm opt}$ = 31\farcs5) and
adjacent regions is less pronounced, but is still significant till the
radial distance of $R \sim$34\arcsec.

We fitted the $V,R$ SBPs by the `minimal' exponential disks, using for
this only the ranges in radial coordinates where the contribution
of the superimposed emission is minimal/absent, namely
$20\arcsec\ < R < 40\arcsec$\  and $60\arcsec\ < R < 70\arcsec$.
The derived scalelengths of the disks in $V$ and $R$-bands do not differ
significantly within their uncertainties. Their mean value is
$<$$\alpha$$>$=15\farcs3$\pm$0\farcs3, corresponding to the linear value of
482$\pm$10 pc.
We therefore accept the hypothesis that the underlying light is described by
the unique exponential disk.
The central surface brightnesses of the mean fitted disk are $\mu^{0}_{V}$ =
22.56 mag arcsec$^{-2}$ and $\mu^{0}_{R}$ = 22.34 mag arcsec$^{-2}$ with
a total error of $\sim$0\fm06.
The $(V-R)^{0}$ colour radial profile along with the derived disk colour
(solid line) are shown in the bottom panel of Fig. \ref{ProfVR}.

Total magnitudes of DDO~68 derived by the integration within a polygon mask
including all galaxy light, with the removal of foreground stars and
background galaxies, are as follows:
$V$=14.46$\pm$0.05, $R$=14.20$\pm$0.05, with the resulting
$(V-R)_{\rm total}$=0.26$\pm$0.07.
The relatively large errors appear
from the uncertainties of the coefficients of the photometrical system, built
on observations of the calibration field near QSO FBQ 0951+263.
The total flux of DDO~68 in the H$\alpha$-line, derived within the same
polygon mask,
is (1.73$\pm$0.09)$\times$10$^{-13}$ erg~cm$^{-2}$~s$^{-1}$, which in turn
corresponds to a total H$\alpha$ luminosity of 0.88$\times$10$^{39}$
erg~s$^{-1}$. We did not transfer this value to the traditional star
formation rate (SFR), since for young starbursts the SFR derived through the
commonly used formulas is highly uncertain (see Weilbacher \&
Fritze-v.Alvensleben (\cite{WvA}). However, we compare this and
other global parameters with those in I~Zw~18 and SBS 0335--052 in the
discussion.

Since the `minimal' disk describes the widespread underlying stellar
population over the whole galaxy volume, its colours can be considered
as indicating the oldest stellar population in the galaxy. This colour
\mbox{$(V-R)_{\rm disk}$=0\fm22}, with the rms error of 0\fm07 due to the
photometric system uncertainties. A small correction of $\sim$0\fm012
for the Galaxy extinction should also be applied before comparison with the
evolutionary tracks of PEGASE.2 models (Fioc \& Rocca-Volmerange
\cite{PEGASE,PEGASE2}). We took the models for the metallicity
of $z$=0.0004, the nearest to that of DDO~68.
The Salpeter IMF is accepted with M$_{\rm low}$ and M$_{\rm up}$ of
0.1 M\sunn\ and 120 M\sunn, respectively. Various modes of SF
(instantaneous and continuous) result in different age constraints on the
oldest visible stellar population.

Taken at face value, the disk colour ($(V-R)^{0}$ $\sim$0.21) with the
cited error of $\sigma_{(V-R)}=$0\fm07 corresponds to either
0.65--1.8--4.5 Gyr-old (lower and upper limits correspond to --1$\sigma$
and +1$\sigma$)
stellar clusters for the continuous SF with constant SFR or to
0.11--0.12--1.0 Gyr-old cluster for the instantaneous starburst.

\subsection{Properties of the young star cluster in Knot 5}
\label{Knot5}

The young star cluster (YSC) related to Knot 5 is one of a few known
with such low metallicity that can be directly studied in great detail.
The others include, e.g., some stellar clusters containing WR stars in
I~Zw~18 (Brown et al. \cite{Brown02}) or unresolved with HST and VLT
young clusters in SBS 0335--052~E (Thuan et al. \cite{TIL97}, Plante \&
Savage \cite{Plante02}).
Apart from the difference in metallicity by a factor of $\sim$30, it
resembles the young super star cluster in the bubble complex of NGC 6946
very well (Elmegreen et al. \cite{Elmegreen00},
Efremov et al. \cite{Efremov02}, Larsen et al. \cite{Larsen01}), which is
situated at a similar distance of $\sim$6 Mpc. The latter was successfully
observed with the HST and Keck HIRES to resolve its stellar content and
velocity dispersion and to address its evolutionary status. While for the
YSC in DDO 68, such observations are feasible in the future as well, here we
summarise the properties that emerge from  current observational data.

We can estimate the characteristic size of the Knot 5 cluster by
assuming its intrinsic intensity distribution to be Gaussian with the
respective FWHM$_{\rm YSC}$.
Subtracting (quadratically) the PSF FWHM (0\farcs8) from the observed FWHM
(1\farcs12), we obtained the FWHM$_{\rm YSC}$ intrinsic diameter of
$\sim$0\farcs8.
The respective linear diameter of this cluster is 25 pc. For comparison,
compact young clusters in SBS 0335--052~E, unresolved on the HST images
(the PSF FWHM=0\farcs2) have diameters less than 50 pc (Thuan et al.
\cite{TIL97}).
The further analogy of the DDO 68 young star cluster with the mentioned
above in
NGC 6946 extends to their linear size. As described by Larsen et al.
(\cite{Larsen01}), this cluster has a core with a radius of 1.3 pc,
surrounded by an extended halo with the power-law luminosity profile. The
halo half-light radius is 13 pc, close to half of the FWHM of the light
profile of Knot 5 in DDO~68. With the age of 12-15 Myr (Efremov et al.
\cite{Efremov02}), the young star cluster in NGC~6946 is, however,
significantly younger.

The total $V$=19\fm40 of the YSC in DDO~68 corresponds to the
absolute magnitude \mbox{M$_{V}^{0}$=--9.66}. By comparing this
M$_{V}^{0}$ with that from
the PEGASE.2 model for instantaneous starbursts of the metallicity $z=0.0004$
cluster with the Salpeter IMF for the age T=22.5 Myr (M$_{V}$=2.30 per
1~M\sunn), we derived the mass of the young cluster of
M$_{\rm YSC}$=6.0$\times$10$^4$~M\sunn.
This cluster is the most massive in the `Northern ring'. The brightest among
the others, Knot 1, with  $V$ $\sim$20\fm3, has a star cluster mass of
4.4$\times$10$^3$~M\sunn.

The measured colour of Knot 5  $(V-R)^{0}$ = 0\fm11$\pm$0\fm01$\pm$0\fm07,
where the second component of the error comes from the photometric system,
is redder by 0\fm09 than  predicted by the PEGASE.2 model for its age
($(V-R)$ = 0\fm02).
This is most probably related to some systematics in the zero-points of
the constructed photometric system (see Sect. \ref{correct}).

\subsection{Colour correction and improved age estimates}
\label{correct}

The main uncertainties of the derived $(V-R)$ colours of the underlying disk
and of the estimate of the age of the oldest visible stellar
population are related to the zero-points uncertainty of the constructed
photometrical system ($\sigma_{\rm (V-R)}$ = 0\fm07). Having some independent
observational data for DDO~68,  we can now diminish this effect.
For this we used two different approaches.

First, we used the spectral information (EWs of Balmer absorption and emission
lines) for Knot 5 - the young compact star cluster in the northern `ring',
which  implies its age of $\sim$22.5 Myr. Then, as shown at the end of
the previous section, the observed \mbox{$(V-R)$} colour is 0\fm09 redder than
predicted by the PEGASE.2 models for its respective age. This implies the
presence of the additive small shift of 0\fm09 in the $(V-R)$ zero-point
due to the photometric system uncertainties.

Second, there is an independent indication of the reality of the $(V-R)$
colour shift found above. This comes from the comparison of our measured
$(V-R)_{\rm total}$ = 0.26 and the colour $(B-V)_{\rm total}$ = 0.17
from Karachentsev et al. (\cite{NGC}) in Table \ref{tab:Main_par}, with
\mbox{$(B-R)_{\rm total}$} = 0.32$\pm$0.03 from Hopp \& Schulte-Ladbeck
(\cite{HS95}).
Combining the former, our value with the cited $(B-V)_{\rm total}$
led to $(B-R)_{\rm total}$ = 0\fm43, while using the above correction
brought this parameter to value of 0\fm34, significantly more consistent
with that of Hopp \& Schulte-Ladbeck.

Now we apply this correction of the $V-R$ zero-point and the respective
reduction of its uncertainty roughly to 0\fm04 from comparison
with the Hopp and Schulte-Ladbeck $(B-R)_{\rm total}$
to the colours of the underlying population.
The corrected $(V-R)^{0}$ colour of the underlying disk
is 0\fm12$\pm$0\fm04. This in turn
corresponds (in the frame of the same PEGASE.2 models as above) to the ages
of 200 -- 450 -- 900 Myr for the case of continuous SF with constant SFR, or
to the ages of 100 -- 105 -- 115 Myr for the case of instantaneous starburst.
The lower and upper limits correspond, as above, to $-$1$\sigma$ and
+1$\sigma$ in the $(V-R)$ colour.

Applying the above correction of $(V-R)$, we somewhat improve the
accuracy of the derived ages of the old population. However, due to the well
known degeneracy of the colour evolutionary tracks for different SF laws,
the estimated ages are still uncertain by a factor of \mbox{$\sim$(2--8)},
if based only on the optical colours of unresolved stars, depending on the
accepted SF law.  Besides, the use of the other optical-NIR colours is
important to check the results suggested by the study of only $V-R$  colour.

The most reliable age estimates can be obtained from the colour-magnitude
diagrams (CMD) of resolved stars. If the galaxy is indeed young, no RGB
stars will be seen in the CMD, as in the case of I~Zw~18 (Izotov
\& Thuan \cite{IZw18CMD}). For the distance modulus of DDO~68 equal to 29.0
(for $D=6.5$ Mpc), stars on the tip of RGB should have $V$=25\fm0
and $I$=23\fm8.

\section{Discussion and Conclusions}
\label{discussion}

\subsection{Properties of DDO~68 in comparison to I~Zw~18}

The weighted mean 12+$\log$(O/H)=7.21$\pm$0.03 for DDO~68
is very close to that in I~Zw~18. Besides, DDO~68 is close
to I~Zw~18 on several other parameters. In particular, its absolute blue
magnitude $M_{B}$ is only 0\fm56 fainter.
Thus, both objects are very close on the L--Z diagram and,
together with SBS~0335--052 E and W, comprise a group of galaxies
significantly deviating from the general L--Z relation for BCGs (see,
e.g., Pustilnik et al. \cite{Kiel}).  Such
deviations are also noticed for several luminous BCGs with 12+$\log$(O/H)
$\sim$8.0 (e.g., Kunth \& \"Ostlin \cite{Kunth2000}), which are old galaxies,
so the strong deviation from the general L--Z relation is not an exclusive
property of the most metal-deficient BCGs. However, the objects of this small
group
deviate as well on the L--Z diagram
from the region where the majority of known XMD galaxies are situated (see
Kniazev et al. \cite{Kniazev03}).
Based on this large deviation, Kniazev et al. (\cite{Kniazev03})
suggest that the small XMD subgroup of I~Zw~18 and its `cousins'  are
truly young galaxies. A similar conclusion was formulated for the same
small  subgroup of XMD galaxies by Guseva et al. (\cite{SBS1415}), based
on the very blue $(V-I)$ colours of their LSB components.

The M(\ion{H}{i})/L$_{\rm B}$ ratio for DDO~68 is also sufficiently high
(1.85), and is a factor of $\sim$1.4 higher than  for
I~Zw~18. Its total \ion{H}{i} mass (2.6$\times$10$^{8}$ M\sunn) is
a factor of 1.6 higher than the \ion{H}{i} mass of I~Zw~18.
Its integrated colour ($B-V$)$_{\rm tot}^{0}$=0\fm17
(Makarova \& Karachentsev \cite{Makarova98})  is  close to that
of I~Zw~18  (0\fm12). Some other properties of these two
galaxies are collected in Table \ref{tab:Main_par} for comparison.
Direct comparison of the current SFR (based on the H$\alpha$-luminosity)
of DDO~68 and other similar galaxies is not meaningful, since this SFR changes
significantly on a short timescale. However the specific
values of H$\alpha$-luminosity related either to the blue luminosity or to
the total \ion{H}{i} mass are of interest.
For further comparison, we take F(H$\alpha$) of I~Zw~18 from Gil de Paz et al.
(\cite{Gil03}).

In DDO~68 these parameters are significantly lower than for I~Zw~18,
indicating a lower current production of the ionizing radiation.
Namely, L(H$\alpha$)/L$_{\rm B}$
is $\sim$0.005 for DDO~68 and 0.035 - for I~Zw~18. Similar,
L(H$\alpha$)/M(\ion{H}{i}) is 0.0016 (in units $(M/L_{\rm B})$\sunn)
for DDO~68 and 0.028 -- for I~Zw~18. Interestingly,  on these parameters
DDO~68 is closer to SBS 0335--052~W, where these ratios are
L(H$\alpha$)/L$_{\rm B}$ = 0.02 and L(H$\alpha$)/M(\ion{H}{i}) = 0.0027
(H$\alpha$-flux is from Pustilnik et al. \cite{BTA}). It was noticed that
based on its properties, SBS 0335--052~W is not a typical blue compact galaxy,
but instead resembles dIrr, which based on the strength of the current
`starburst' is also correct for DDO~68.
The close similarity of principal observational properties of DDO~68
and I~Zw~18 suggests that DDO~68 may also  be a young galaxy,
that is it contains no stellar population with ages more than 1 Gyr.

\subsection{Optical and \ion{H}{i} morphology and their implications}

The strongly disturbed optical morphology of DDO~68 is difficult to
understand since the galaxy is rather distant from any potential disturber.
The luminous
galaxies are at least at $\sim$2.0 Mpc (NGC 2683 with M$_{B}$=--20.4 at
D=7.7 Mpc) and $\sim$2.5 Mpc (NGC 2903 with M$_{B}$=--21.0 at D=8.9
Mpc, Karachentsev et al. \cite{NGC}). The nearest known galaxy, a dwarf
UGC 5427, with $M_{B}$ = --14.5 at $\sim$200 kpc in
projection and with very close radial velocity, is too low-mass object to
affect DDO~68.
The neutral gas morphology and kinematics, as revealed by \ion{H}{i}
maps with 13\farcs5 angular resolution by Stil \& Israel
(\cite{Stil02a, Stil02b}), are also well disturbed with several maxima of
density and with several holes with linear sizes of the order of 1 kpc.

One could consider two options to understand the optical morphology of DDO~68.
The first is related to the strong disturbance by some separate, massive,
poorly visible (or optically invisible) object. This could be a very low
surface brightness galaxy, or an intergalactic \ion{H}{i} cloud, similar to
the case of HI 1225+01 (Salzer et al. \cite{Salzer91}, Chengalur et al.
\cite{Chengalur95}).
The second option is related to an advanced merger, when two merging objects
are already  not seen as separate entities. However, the morphology
of the joint material clearly indicates outflows/plumes as the
characteristic features created by  strong interaction. The importance
of the merger channel for starbursts in BCGs was emphasised, e.g.,
by \"Ostlin et al. (\cite{Ostlin01}).
Of course, the combined option is also possible: for example, a well-advanced
merger with an intergalactic \ion{H}{i} cloud.

To distinguish the above options, the available \ion{H}{i} maps are very
useful. We also used overlays of the \ion{H}{i} density map on the DDO~68
$I$-band and H$\alpha$ images, along with detailed data on \ion{H}{i}
velocity dispersion in the Ph.D. thesis of Stil (\cite{Stil99}).
Analysis of all these data leads to the next conclusions. First,
there is no indication of the presence of any separate massive \ion{H}{i}
object in the vicinity of DDO~68.  Second, there is the large asymmetry of
\ion{H}{i} density distribution relative to the optical body of the galaxy.
The \ion{H}{i} density map can be divided conditionally into two curved
chains of dense matter stretching roughly from North to South, with the
typical
separation between the chains of 0\farcm5--1\farcm0 (1--2 kpc). The optical
body, including the Northern `ring' and the Southern `tail', coincides
roughly with the Eastern \ion{H}{i}
chain, while practically no optical counterparts are seen for the Western
\ion{H}{i} chain (at least, on the SB level, comparable with the optical
light seen along the Eastern \ion{H}{i} chain).

The large asymmetry between the distributions of \ion{H}{i} mass and
the optical light suggests that the Western part of the \ion{H}{i} body is
unrelated directly to the optical galaxy and represents some external
\ion{H}{i} cloud in the process of merging with a gas-rich massive object,
in which some star formation has already occurred and is still taking place.
The elevated velocity dispersion
in the four densest regions (by a factor of 2 to 3 relative to the undisturbed
regions, which have $\sigma_{v}$=6--9~\kms), all them being displaced to the
Eastern half of the whole \ion{H}{i} cloud, gives additional evidence of
strong gas agitation due to the ongoing merging, as suggested by
Elmegreen et al. (\cite{Elmegreen93}). If this hypothesis is correct,
DDO~68 can represent the well-advanced stage of merging observed in the system
HI 1225+01. The metallicity of gas in the two merging components can
differ substantially, so the complex process of the gas mixing can create
significant spatial inhomogeneities in the ISM metal content of DDO~68.

While the WSRT DDO~68  \ion{H}{i} map does not show the extended tidal
tails characteristic of many mergers, this is not an argument
against the merger case. However, the tails are not likely likely to be seen
because the available maps are relatively shallow,  with the
lowest \ion{H}{i} density contour of 5$\times$10$^{20}$ atom~cm$^{-2}$.
The extended tidal tails in the well-advanced mergers like those
in NGC~7252, NGC~3921, NGC~3526 (e.g., Hibbard et al. \cite{gallery}), are
visible at the
\ion{H}{i} column densities of (one -- a few) 10$^{19}$ atom~cm$^{-2}$.
Thus, in the frame of the merger hypothesis, we predict that
the deeper \ion{H}{i} mapping of DDO~68 will uncover the extended gas
tidal tails.

We also mention a more exotic variant of the first option, when the
strong interaction of a gas-rich dwarf took place with a completely dark
galaxy, which is a massive DM halo with the baryon mass fraction well
below that typical of galaxies (Trentham et al. \cite{Trentham01}). As these
authors suggest, this can be the trigger of starbursts in many
isolated BCGs. However, to check this hypothesis observationally, they need
to have very clear-cut predictions, since at first approximation it is
difficult to distinguish this case from well-advanced merger.

\begin{table}[h]
\centering{
\caption{Main parameters of DDO~68 and I~Zw~18}
\label{tab:Main_par}
\begin{tabular}{lrr} \hline\hline
Parameter                           & DDO~68$^{1}$      & I~Zw~18              \\ \hline

Type                                &Im/BCD             & BCD                  \\
$B_{\rm tot}$                       &14.60              & 16.12$^{2}$          \\
$(B-V)_{\rm tot}$                   &0.17               & 0.11$^{2}$           \\
$(V-R)_{\rm tot}$                   &0.17$^{3}$         & 0.12$^{2}$           \\
$V_{\rm Hel}$ (km s$^{-1}$)         & 502               &751$^{4}$             \\
$D$(Mpc)                            & 6.5$^{3}$         & 15$^{8}$             \\
$E(B-V)$$^{5}$                      & 0.018             & 0.032                 \\
$M_{B}^{0}$                         &  --14.33          & --14.89              \\
Angular size (\arcsec)$^{\ddag}$    &103$\times$38$^{3}$  & 23$\times$13$^{2}$ \\
Optical size (kpc)                  & 3.2$\times$1.2$^{3}$ & 1.5$\times$0.8~   \\
12+$\log$(O/H)  \                   & 7.21              & 7.17$^{6,7}$         \\
H$\alpha$ flux$^{*}$                & 1.73$^{3}$        & 3.52$^{9}$           \\
H {\sc i} flux$^{**}$               & 26.7              & 2.97$^{4}$           \\
$W_{50}$ (km s$^{-1}$)              & 78                & 49$^{4}$             \\
$W_{20}$ (km s$^{-1}$)              & 85                & 84$^{4}$             \\
$M$(H{\sc i}) (10$^{8}$$M_{\odot}$) & 2.6               & 1.6                  \\
$M$(H{\sc i})/$L_{B}$$^{***}$       & 1.85              & 1.29                 \\
\hline\hline
\multicolumn{3}{l}{$B_{\rm tot}$ -- total blue magnitude; $M_{B}^{0}$ -- absolute blue mag.} \\
\multicolumn{3}{l}{$L_{B}$ -- total blue luminosity. $^{*}$\ Units of 10$^{-13}$ erg~s$^{-1}$;} \\
\multicolumn{3}{l}{$^{**}$\ Units of Jy km s$^{-1}$; $^{***}$ In units of ($M$/$L_{B}$)$_{\odot}$.}  \\
\multicolumn{3}{l}{$^{\ddag}$\ $a \times b$ at the surface brightness $\mu_{B}$=25 mag arcsec$^{-2}$.} \\
\multicolumn{3}{l}{{\bf References}: $^1$DDO~68 parameters without reference } \\
\multicolumn{3}{l}{are from Karachentsev et al. (\cite{NGC}); $^{2}$Papaderos } \\
\multicolumn{3}{l}{et al. (\cite{Papa02}); $^{3}$ -- this paper; $^{4}$van Zee et al. (\cite{vanZee98};} \\
\multicolumn{3}{l}{$^{5}$Schlegel et al. (\cite{Schlegel98}); $^{6}$Izotov et al. (\cite{Izotov99}); } \\
\multicolumn{3}{l}{$^{7}$Kniazev et al. (\cite{Kniazev03}); $^{8}$Izotov \& Thuan (\cite{IZw18CMD});} \\
\multicolumn{3}{l}{$^{9}$Gil de Paz et al. (\cite{Gil03})} \\
\end{tabular}
 }
\end{table}

\subsection{Chemical evolution and environment}
\label{environ}

In the frame of Cold Dark Matter cosmology the
galaxies formed in the regions of the very low density (voids) should be
of lower mass and retarded in their formation epoch in comparison to the more
massive galaxies from the average and the higher density regions (e.g.,
Gottl\"ober et al. \cite{Gott03}).
Besides, in the frame of the widely accepted hierarchical galaxy formation
scenario, their life cycles can differ quite a lot from the more common
galaxies. Namely, for a selected galaxy, the probability of interactions and
mergers during the cosmological time is scaled down in voids with the galaxy
density as $\rho_{\rm galaxy}^{+1}$.
Therefore, one could expect that some (small) fraction of void galaxies can
survive in their nascent state, avoiding any significant interaction.

It is probably not by chance that one of the most metal-poor  galaxies,
DDO~68, is found in the very low-density `normal' galaxy environment.
Another XMD galaxy HS 0822+3542 is situated near the centre of the same
Lynx-Cancer void (Pustilnik et al. \cite{P_SAO0822}). At least one more
dwarf galaxy with very low metallicity, KISSB~23 = KUG~0937+298
(12+$\log$(O/H)=7.65, Lee et al. \cite{Lee04}) is also situated within the
same void, only at $\sim$650 kpc from DDO~68.

The nearest neighbour of DDO~68 is UGC~5427 (Sdm, B$\sim$14\fm9) at the
angular distance of $\sim$1.8\degr, corresponding to $\sim$200 kpc
in projection.
Its radial velocity is only 7~\kms\ lower than that of DDO~68.
Karachentsev et al. (\cite{NGC}) accept distances  for DDO~68 and UGC~5427
derived by the method of the brightest stars (accuracies of 25\%), of 5.9
and 7.1 Mpc, respectively. We suggest that both galaxies belong to the
same `filament' and the distance in between  is on the order of their
projected distance. Therefore, we accept their common radial distance as a
mean of their individual determinations, consistent each to other:
$D_{\rm mean}=6.5$ Mpc.

As discussed by Pustilnik et al. (\cite{BTA}), a substantial
fraction of XMD galaxies are found either at the outskirts of galaxy
aggregates (e.g., SBS~0335--052 E,W and Dw 1225+0152), or in voids
(e.g., HS 0822+3542, HS 2236+1344, HS 0837+4717). While the statistics
and properties of void galaxy population are still not well known, it is
assumed from general consideration that the significantly reduced
frequency of galaxy interactions in voids provides the most favourable
conditions to allow for the stablest gas protogalaxies to survive as
purely gas objects.
If such very stable protogalaxies do exist in voids, they can be
detected either as purely gas bodies through the blind \ion{H}{i} surveys
or through the Ly-$\alpha$ absorption of a background UV-bright object
(e.g., Manning \cite{Manning02}), or even as young galaxies,
if they recently have experienced  the strong gravitational disturbance due to
close collision with sufficiently massive objects.

Therefore,  on one hand,   it is not surprising that many `void' XMD
galaxies studied by us (a paper in preparation) show clear evidence of
interactions and merging.
This can be a kind of selection effect for the superstable objects, when
starbursts in unevolved galaxies occur only due to the strong external
disturbance.
The case of DDO~68 is additional evidence for the common mode of such
starbursts in the XMD galaxies situated in the very low density environment.

On the other hand, as explained by Kunth \& Sargent (\cite{KS86}),
even the first starburst in such a retarded object, due to the effective
self-pollution in the young \ion{H}{ii} regions, will be observed with
a heavy element content comparable to Z $\sim$Z\sunn/30 for the starburst
ages larger than 3 Myr.

\subsection{Conclusions}
\label{Conclusions}

Summarising the observational data and the discussion above, we have drawn
the following conclusions:
\begin{itemize}
\item The oxygen abundances in three \ion{H}{ii} regions at the northern edge
    of the dwarf galaxy DDO~68 are close one to another with the weighted
    mean of 12+$\log$(O/H)=7.21$\pm$0.03 dex.
\item  DDO~68 is one of the three galaxies (I~Zw~18 and SBS 0335--052~W)
    with the lowest ISM metallicity, and the nearest galaxy with the same
    low metallicity (twice closer than I Zw 18).
\item The deep photometry of DDO~68 indicates that its the reddest colours
      outside the central bright region and the regions of current/recent SF
      (where no nebular emission is seen) are very blue ($(V-R)^{0}$=0.12
	$\pm$0\fm04). Their comparison with the PEGASE.2 models for the
      evolving stellar clusters (with metallicity of $z=0.0004$) implies
      that the oldest visible stars contributing to the light of the LSB
      underlying disk have ages either of 100--115 Myr. in the case of
      instantaneous starburst, or of 200--900 Myr, in the case of continuous
      star formation with the constant SFR. Thus, $V-R$ data indicate
      the possible youth of DDO~68. The data on other colours, as well on
      CMDs, are necessary, however, to check and strengthen this conclusion.
\item The young star clusters (related, in particular to Knots 1 and 5
       in the `Northern ring') with the masses of
       (0.4--6.0)$\times$10$^{4}$~M\sunn\ are the nearest among all known
       such objects with the lowest metallicity. They are probably the best
       known clusters for which the direct spectroscopic study of
       {\it individual massive} stars with  Z $\sim$Z\sunn/30 will be
       possible with the future giant ground-based and space telescopes.
       Study of their structure and dynamics is currently feasible with
       the HST and the high resolution spectroscopy at large telescopes.
\item The optical morphology of DDO~68 indicates recent interaction
      despite no disturbing galaxy is evident in its surrounding. The
      most probable interpretation of this galaxy's properties is a
      well-advanced merging of two gas-rich objects.
\item  DDO~68 is situated in the region with the low density of luminous
      galaxies near the rim of a nearby small Lynx-Cancer void. Apart from
this galaxy, two more XMD dwarfs have been  discovered in this void:
      HS 0822+3542 and KISSB~23. Several other XMD galaxies are found in
      other voids. This is consistent with theoretical expectations that
      dwarfs in the regions with the very low density of luminous galaxies
     (and the very reduced density of low-mass galaxies) may form/evolve with
      significant delay. This opens a new direction in searching for
      local young  galaxies.
\end{itemize}

\begin{acknowledgements}

The authors are pleased to thank V.~Afanasiev and A.~Moiseev for
the help in the organization and preparation of observations with SCORPIO,
Y.~Izotov for useful comments, and J.~Stil for kindly providing his Ph.D.
thesis. We appreciate the criticism and suggestions of the
referee, D.~Kunth, which helped to improve the paper.
We acknowledge the partial support from Russian state program
"Astronomy". This research made use of the NASA/IPAC Extragalactic
Database (NED), which is operated by the Jet Propulsion Laboratory,
California Institute of Technology, under contract with the National
Aeronautics and Space Administration. Use of the Digitized Sky Survey
(DSS-II) is gratefully acknowledged.

\end{acknowledgements}

\renewcommand{\baselinestretch}{0.5}

\end{document}